\documentclass{vgtc}                          % final (conference style)

\graphicspath{{figures/}{pictures/}{images/}{./}} % where to search for the images

\usepackage{times}                     % we use Times as the main font
\usepackage{tabu}                      % only used for the table example
\usepackage{booktabs}                  % only used for the table example
\usepackage{lipsum}                    % used to generate placeholder text
\usepackage{mwe}                       % used to generate placeholder figures
\usepackage{amsmath}
\usepackage{siunitx}
\usepackage{amsmath}
\usepackage{algorithm}
\usepackage{algpseudocode}

\usepackage{comment}

\usepackage{geometry}
\usepackage{array} % For wrapping text in table cells
\usepackage{mathptmx}                  % use matching math font

\onlineid{1178}

\vgtccategory{Research}

\vgtcinsertpkg

\title{Offsetting Perceptual Bias in Visual Clustering: The Role of Point Size Adjustment in Variable Display Sizes}

\author{Taehyun Yang\thanks{e-mail: 0705danny@snu.ac.kr}, Hyeon Jeon\thanks{e-mail: hj@hcil.snu.ac.kr}, Jinwook Seo\thanks{e-mail: jseo@snu.ac.kr}}
\abstract{
Scatterplots are frequently shared across different displays in collaborative and communicative visual analytics. 
However, variations in displays diversify scatterplot sizes.
Such variations can influence the perception of clustering patterns, introducing potential biases leading to misinterpretations in cluster analysis. 
In this research, we explore how scatterplot size affects cluster assignment and investigate how we can offset such bias.
We first conduct a controlled study asking participants to perform visual clustering on scatterplots of varying sizes. We found that changes in scatterplot size significantly alter cluster perception in three key features.
In our subsequent experiment, we examine how adjusting point sizes can mitigate this bias. As a result, we verify that adjusting point size can effectively counteract the perceptual biases caused by varying scatterplot sizes.
We wrap up our research by discussing the necessity and applicability of our findings in real-world applications.
}


\begin{document}

%% The ``\maketitle'' command must be the first command after the
%% ``\begin{document}'' command. It prepares and prints the title block.

%% the only exception to this rule is the \firstsection command

\maketitle

%% \section{Introduction} %for journal use above \firstsection{..} instead

% Introduction.tex
\section{Introduction}
%Scatterplots are widely-used visualization technique to represent multi-dimensional data on a two-dimensional plane

%first sentence is wayy too general multiple size 이 첫 문단 이 되어야하고 두번째 문단이 이문제를 scatterplot 에서 적용.

Scatterplot is one of the most widely used visual idioms. Using scatterplots, users can readily grasp complex patterns and trends within datasets~\cite{Ziefle1999}. Due to its importance, several research works investigated how visual perception is affected by the design of scatterplots \cite{Abbas19,Jeon24,Hong22}. 

Scatterplots are commonly shared across different displays with various sizes and resolutions for collaborative visual analytics and data communication~\cite{NGUYEN20201,Sarikaya2018}. For example, instructors often use scatterplots to present data to students. Similarly, scatterplots can be used to share data analytics results with stakeholders or domain experts.
Here, to present scatterplots in various displays, visualization designers or analysts commonly scale up the scatterplot or point sizes to fit into the new screen size and resolution~\cite{Cleveland1984The}. However, such naive scaling incurs biases in perceiving data patterns~\cite{Wei_2020}, leading to inconsistencies in data analysis. 

 \begin{figure}
    \centering
    \includegraphics[width=\linewidth]{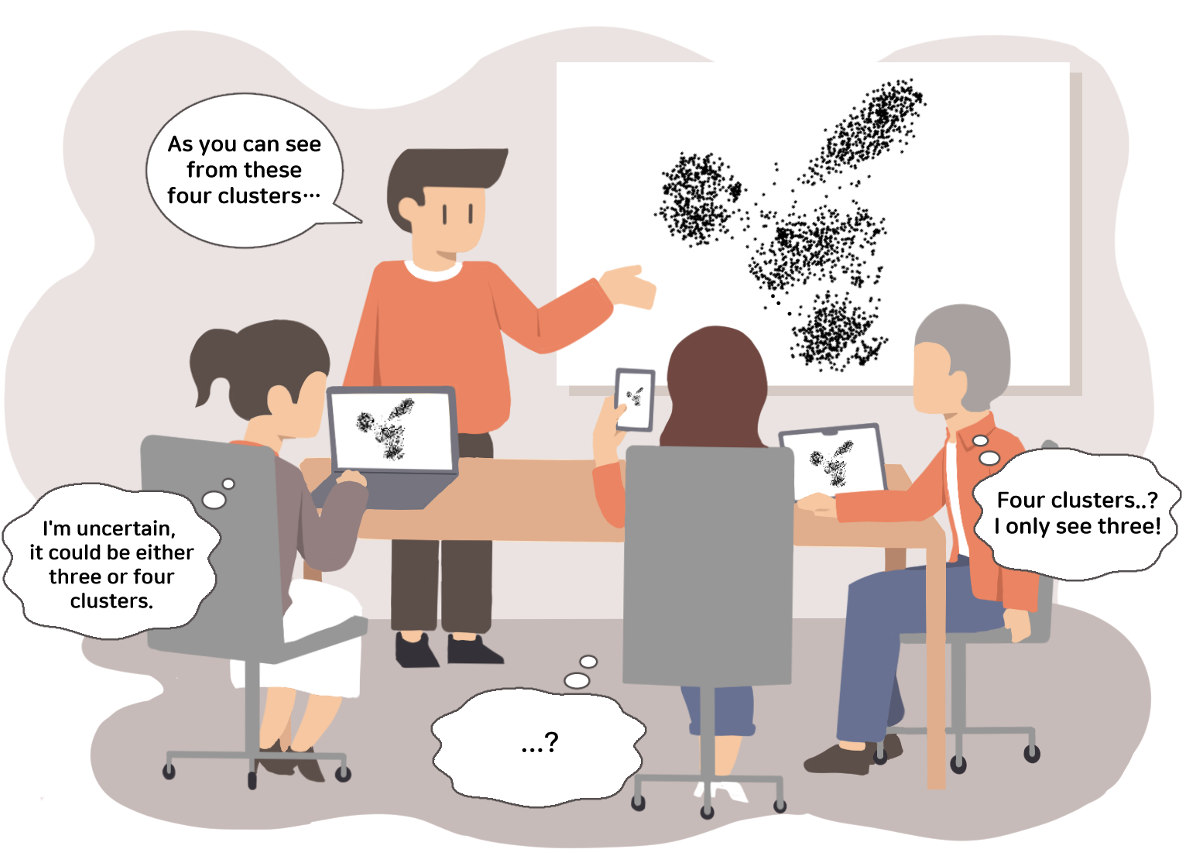}
    \caption{Illustration of our research motivation: the perceptual bias in visual clustering due to scatterplot sizes. In communicative or collaborative analytics environments (e.g., classrooms), users often view the same scatterplots on different displays. These displays present scatterplots in various sizes, which can lead to different interpretations in visual clustering analysis.
    }
    \label{fig:smallscatter}
\end{figure}

Previous research works have investigated how human perception is influenced by different scatterplot features, including point color~\cite{Szafir18,Ware04}, mark types~\cite{Lewandowsky89,Heer10}, and aspect ratio~\cite{Cleveland88,Heer06}. These findings guide designers and analysts in optimizing scatterplot design to meet specific needs. In terms of point size and scatterplot size, Wei et al. \cite{Wei_2020} identified potential biases in the visual task of assessing (1) which scatterplot has more points (numerosity), (2) which scatterplot shows stronger correlation (correlation), (3) which scatterplot exhibit more obvious separation between two different colored classes (class separation). Furthermore, the study also stated that biases in these three tasks could be partially corrected by changing the point size. However, to the best of our knowledge, no prior study has quantitatively investigated how scatterplot size affects \textit{cluster perception} or \textit{visual clustering} and whether adjusting point sizes can help maintain consistent clustering perception across varying monochrome scatterplot sizes.

Our research fills this research gap by providing (1) empirical evidence to identify the perceptual biases that occur when scatterplot sizes change and (2) verification that point sizes can be adjusted to alleviate the perceptual biases identified in (1). This enables a more accurate and consistent cluster perception across various display environments, which is essential for reliable data analysis and decision-making processes. 
% This contribution is important for developing automated tools that adjust visual parameters dynamically, ensuring accurate data interpretation regardless of display conditions.

To do so, we first conduct a preliminary study investigating how visual clustering results vary by scatterplot size. 
The results indicate that the cluster perception is significantly affected by scatterplot size.
Larger scatterplots showed a smaller number of clusters, more points within each cluster, and a higher relative area for each cluster.
% For our preliminary study, we meticulously select 20 monochrome scatterplots from real-world data, with sizes randomly assigned as large or small, and task  15 users with lassoing clusters they perceived. 
Subsequently, in our main study, we systematically adjust point sizes across 20 randomly sized monochrome scatterplots and measure how these modifications affect participants' cluster perception. 
As a result, we found that adjusting point sizes can substantially mitigate perceptual bias caused by varying scatterplot size.
% /

% The results indicate significant differences in the granularity of cluster assignments between these two sizes. 
% Finally, based on the study results, we conduct an experiment that verifies that incremental changes of point size can e/ffectively offset perceptual bias that occurs when scatterplot size is changed.

% changes in cluster number and alterations in point size and scatterplot size for specific scatterplots. 
% We then correlated the results of the cluster granularity model with insights gathered from subjective questionnaires. 

To demonstrate the effectiveness of the research, we discuss the potential future works and applications of our findings.
Specifically, our findings can ground the development of a mathematical model to calibrate each scatterplot to the display size of different devices, ensuring that all participants have a consistent view of the data. Such model can also be integrated into software tools used by data analysts and scientists, allowing for automatic adjustments of visual encodings to ensure consistent representation of data clusters regardless of window sizes.

In summary, our study provides the following contributions:
\begin{enumerate}
    \item Initial findings on how cluster assignments vary between large and small scatterplots.
    \item Verification that adjustments to scatterplot point sizes effectively counteract perceptual biases caused by scatterplot dimensions.
    \item  Proposal of potential applications of our findings in collaborative environments and responsive visualization.
\end{enumerate}
    % Related_Works.tex
\section{Related Work}

\subsection{Visual Clustering in Scatterplots}

Visual clustering, the task of grouping similar data points visually, is crucial for exploring and understanding various data patterns \cite{Sarikaya2018, Jeon24}. Sarikaya et al. \cite{sarikaya2018design} described visual clustering as a high-level data summarization task defined as the ability to identify groups of similar items. 

Given the importance of visual clustering, several previous works have concentrated on understanding how human visual perception influences clustering in scatterplots \cite{Clevland84,Lothar95,Lewandowsky89}. Xia et al. \cite{xia2021visual} extended these observations by constructing a large dataset of scatterplots used to predict the influence of human visual perception on clustering outcomes. ScatterNet, a deep learning model, captures perceptual similarities between scatterplots to emulate human clustering decisions \cite{ma2018scatternet}. Furthermore, ClustMe used visual quality measures (VQM) to model human judgments to rank scatterplots \cite{abbas2019clustme}. Recently, CLAMS introduced a VQM that measures ambiguity in cluster perception while performing visual clustering \cite{Jeon24}. Comprehensively, Sedlmair et al. \cite{Sedlmair12} proposed a taxonomy which showed scatterplot features such as scale, shape, and density impact perceived cluster separation differently from preexisting VQM.
% Studies have also investigated the human perception of scatterplots in particular scenarios altering scatterplot features such as the color of points, aspect ratio, dimensions, and even animations \cite{Szafir18,Ware04,Heer10,Cleveland88}. 
However, existing studies have mainly focused on visual clustering when scatterplot and point sizes are fixed. In our research, we aim to investigate how visual clustering changes when scatterplot and point sizes are varied.

\subsection{Effect of Scatterplot Size and Cluster Perception}

The necessity for analyzing data on different displays have increased with the development of display devices \cite{Jakobsen2013Interactive}. e.g., tabletop displays or smartphones. Liao et al. \cite{Liao2018Cluster-Based} introduced a cluster-based technique to improve scatterplot analysis and visualization on various displays. Recently, Christopher et al. \cite{Andrews11} noted that display sizes strongly affect the usability of scatterplots. Researchers have further studied the effect of display sizes on design decisions of scatterplots, such as interactions on mobile phones, clustering perception on large displays, and data visualization in 3d environments \cite{Kraus2020}.

% Furthermore, advances in display technology create new scenarios for explorations in the presentation of visualization tools. 
% The increasing diversity in display technologies has necessitated adaptations in how data visualizations like scatterplots are presented across different devices. 

Recently, Wei et al. \cite{Wei_2020} proposed that geometric scaling in scatterplots creates significant perceptual bias. The research highlights that scaling up scatterplots, such as from a laptop screen to a projector, leads to perceptual biases when users compare correlations, estimate the number of points in a class, and assess class separability in two-color scatterplots. 
Wet et al. also showed that changing the point size can partially mitigate biases in these tasks.
However, to the best of our knowledge, there is no prior work that investigates the effect of scatterplot and point sizes on visual clustering. 
To fill this gap, we clarify that scatterplot size also affects visual clustering in scatterplots, and point sizes can be used to mitigate such perceptual bias.

% Although the research showed that changing the point radius can , there are no solutions that guide how point size should be adjusted. In this work, we aim to give foundation to create the solution to solve perceptual bias of examine how perceptual bias, specifically in visual clustering tasks, occurs when the size of monochrome scatterplots changes. This is done by desmontrating that point size can be altered to offset perceptual biases from scatterplot size changes.
% to better understand and address the perceptual challenges posed by scatterplot scaling in varied display environments.

\section{User Study}

We designed our user study in two stages:
 (1) a preliminary study to identify the perceptual biases in visual clustering when scatterplot size is changed; and
(2) a main experiment to test whether changing point size can mitigate the perceptual bias identified in (1). Based on the study results, we verify that point size can be adjusted to offset the perceptual bias caused by scatterplot size

\subsection{Preliminary Study}

\label{sec:preliminary}

\subsubsection{Objectives and Design}

We want to identify the perceptual bias in visual clustering caused by scatterplot size.
To do so, we asked participants to perform visual clustering in \textit{small} (50px) and \textit{large} (350px) monochrome scatterplots. 
These sizes were chosen to simulate scatterplot dimensions as displayed on a mobile device and a large desktop monitor \cite{Wei_2020}.
We asked participants to lasso the clusters they identified in scatterplots, investigating how visual clustering results vary by the size of the scatterplots.

% \noindent
% \textbf{Generating stimuli.}
% We used scatterplot stimuli made by the 

% To identify the perceptual bias in visual clustering, we first explored how scatterplot size affect cluster assignment via a qualitative study. In the study, participants were tasked with completing visual clustering using one of two datasets, each containing a total of 20 scatterplots. Each dataset included 20 monochrome scatterplots, with 10 each in small (50 px) and large (350 px) sizes. These sizes were chosen to simulate scatterplot dimensions as displayed on a mobile device and a large desktop monitor, viewed from the apparatus described below. The first dataset displayed the first half of the scatterplots as a large size and the second half in a smaller size, while the second dataset arranged these sizes in the reverse order.  Participants were randomly assigned to one of two datasets and asked to perform visual clustering using a lasso tool on scatterplots.

% \subsubsection{Dataset Generation and Apparatus}

\noindent
\textbf{Generating stimuli.}
We sampled our scatterplot stimuli from the corpus of scatterplots generated by the ClustMe study \cite{abbas2019clustme}. 
We used ClustMe scatterplots as they are already used to evaluate ClustMe, a VQM that models human perception of cluster patterns. This aligns with our goal of modeling human perception in visual clustering.

We first wanted to exclude scatterplots that are too simple or too complex for visual clustering task over scatterplots \cite{quadri2020modeling}.
Subsequently, we filtered out scatterplots with more than 12 clusters and less than 3 clusters from the corpus. 
We counted the number of clusters using Gaussian Mixture Models while using the Bayesian inference criterion to determine the optimal number of clusters, following Jeon et al. \cite{Jeon24}.
We then randomly sampled 20 scatterplots from the filtered corpus as our final set of stimuli, which we denote as $\mathcal{S} = \{S_1, S_2, \cdots, S_{20}\}$

\noindent
\textbf{Procedure.}
We conducted the experiment in-person, where one experimenter guided the entire study. After signing a consent form, participants underwent a brief tutorial on using the lasso tool. The participants received instructions to use the lasso tool to label any clusters. 
The definition of a cluster was vaguely provided as "groupings of similar data points" \cite{sarikaya2018design}.
The tutorial lasted at most five minutes. 

We then asked participants to lasso clusters in all 20 scatterplots, one at a time. 
For half of the participants, we made $S_1 \sim S_{10}$ and $S_{11} \sim S_{20}$ to be displayed as small and large-size scatterplots, respectively.
For another half, we set the scatterplot size oppositely. 
Note that we fixed all other properties of scatterplots. We displayed data points as black circled dots on a white background and set the size of the point as 8px.

To alleviate the ordering effect, we randomize the order of scatterplots.
To further mitigate potential biases beyond scatterplot size, a blank screen was placed between each scatterplot presentation. \cite{healey2011attention}.

After the trials, we asked participants to fill out a short questionnaire to gather qualitative data on their experience. Each participant's entire session lasted approximately 35 minutes. As a result, we collected 20 X 20 = 400 responses. 

\noindent
\textbf{Participants}
We recruited 20 participants for our preliminary study, comprising students and faculty members from a local university (aged from 20 to 53 years, 26$\pm$ 2.4). None of the participants reported visual or physical disabilities that would impair their ability to view scatterplots or use a keyboard and mouse. Experience with scatterplot analysis varied among participants: 9 had some experience, 4 had extensive experience, and 2 had little or no experience. Participants were compensated upon completion of their tasks.

\begin{figure}[t]
  \centering
  
    \includegraphics[width=\linewidth]{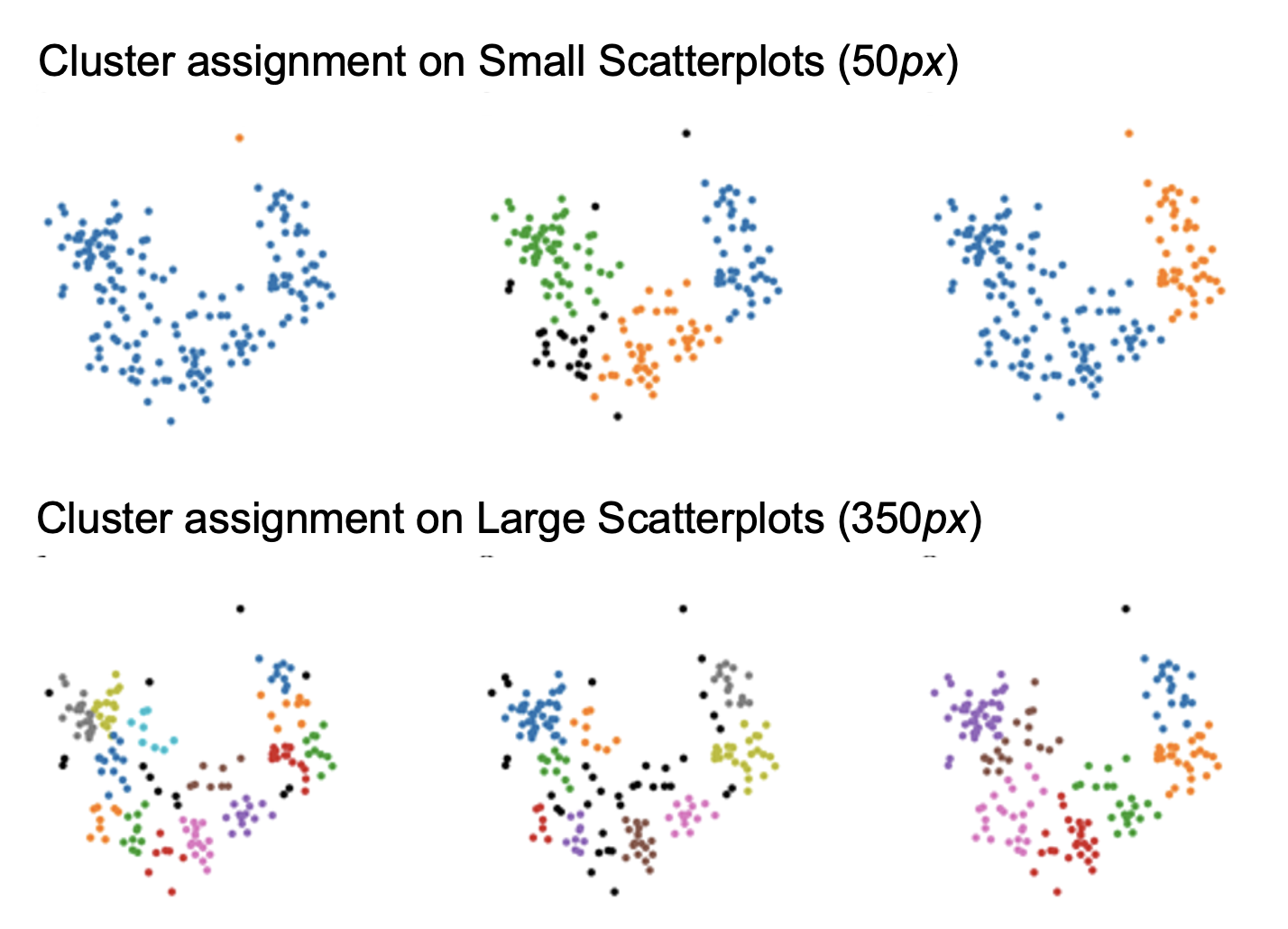}
    \caption{Cluster assignment on small (50px) vs large (350px) scatterplots made by the participants of our user study. Smaller scatterplots lead participants to identify less number of clusters with lower density and bigger cluster area.}
    \label{fig:smallscatter}

\end{figure}

\noindent
\textbf{Apparatus.}
We developed a website in which participants could see scatterplots and complete their tasks with a traditional wireless mouse. Given the significance of consistent scatterplot size and viewing conditions, we conducted the experiment in the same room for all participants. We positioned a 21:9 2K monitor approximately \SI{40}{cm} away from the participants' eyes. Furthermore, we displayed the website on fullscreen so that the scatterplot position and size were consistent. Keyboards were disabled in case any erroneous input caused the website window to be minimized or changed. 

\subsubsection{Results and Discussion}

\label{sec:preresult}
We conducted paired t-tests across 20 datasets to compare measurements between small and large scatterplots. We identify three key changes between these scatterplots: number of clusters, point density, and relative cluster area.

\begin{itemize}
    \item \textbf{Number of Clusters}: Paired t-tests revealed that there was a statistically significant difference in the number of clusters, with larger scatterplots generally exhibiting more clusters.(\( t = 2.45, p < 0.05\));
    \item \textbf{Point Density}: Paired t-test revealed that there was a statistically significant difference in point density, with larger scatterplots exhibiting lower density. Cluster density is calculated by  \( P_{density} = \frac{\text{Number of Points in Clusters}}{\text{Total Points}} \). (\( t = 2.10, p < 0.05 \)).
    \item \textbf{Relative Cluster Area}: Paired t-test revealed that there was a statistically significant difference in relative cluster area, with larger scatterplots exhibiting smaller relative area. This area is calculated by \( A_{relative} = \frac{\text{Area Covered by Clusters}}{\text{Scatterplot Area}} \). (\( t = 1.98, p < 0.05 \)).
\end{itemize}

The results confirm our hypothesis that changes in the size of monochrome scatterplots significantly affect visual clustering results. The increase in the number of clusters identified in larger scatterplots suggests that participants may perceive more distinct groupings. Conversely, the higher point density in clusters observed in smaller scatterplots might reflect a tendency to group data points more tightly. Overall, the cluster assignments in smaller scatterplots were more coarse-grained, resulting in denser and less defined clusters. Meanwhile, cluster assignments in larger scatterplots were more fine-grained, leading to sparser and more defined clusters.

The qualitative feedback from participants demonstrates awareness of perceptual biases, with 14 out of 20 respondents acknowledging that they perceived some form of bias during their clustering activity. Such variation highlights the importance of educational tools or models that can make the perception of clusters consistent across varying scatterplots.

\subsection{Main Experiment}

\subsubsection{Objectives and Design}
We want to explore whether adjusting the point size in scatterplots can counteract perceptual biases caused by changes of scatterplot size. 
We do this by investing in adjusting point sizes to ensure visual clustering to be  maintained consistently against different scatterplot sizes.

To do so, we showed scatterplot stimuli multiple times with different scatterplot and point sizes and checked whether adjusting point sizes and mitigating differences in scatterplot sizes.
This time, we asked participants to count the number of clusters.
We selected a number of clusters among three key differences we identified from the preliminary study (\autoref{sec:preresult}) as it is one of the least difficult tasks in visual clustering, thus reducing the total running time of the experiment. 

Our main assumption here is that increasing point size can mitigate the perceptual bias of visual clustering incurred by the increment of scatterplot size, and vice versa.
Our assumptions build upon Wei's finding \cite{Wei_2020} that the increment/decrement of point size can remove perceptual bias in correlation, density, and class separation tasks due to the increment/decrease of scatterplot size.

The detailed study design is as follows.

\noindent
\textbf{Stimuli.}
We generated stimuli with the same methodology we used for the preliminary study (\autoref{sec:preliminary}). Note that as we randomly sampled scatterplots, the resulting set of stimuli is different from the previous experiment. We again denote these scatterplots as $\mathcal{S} = \{S_1, \cdots, S_{20}\}$.

\noindent
\textbf{Procedure.}
As with the preliminary experiment (\autoref{sec:preliminary}), we conducted the experiment in person. 
After participants signed a consent form, we gave same instruction to participants with the preliminary experiment, except that the task is to count the number of clusters.

We then showed participants a series of scatterplots with different point and scatterplot sizes, and asked participants to count the number of clusters. For each trial, we first randomly sample a single stimulus $S_i$ from $\mathcal{S}$. Then, if $S_i$ has never been shown to the participant, we randomly set the scatterplot size from 50px to 350px.
If it is the second appearance of $S_i$, we randomized the scatterplot size from 50px to 350px but ensured the difference with the scatterplot size of the first appearance to be bigger than 120px. 
This is to identify whether perceptual bias occurred from changing the scatterplot size.
For more than a third appearance of $S_i$, we maintained the scatterplot size from the second appearance but adjusted point size depending on the previous cluster count report. 
If cluster count is reported to be bigger than the initial cluster count, then the point size is increased by 1px. Otherwise, the point size is decreased by 1px. 
If the cluster count is the same as the initial cluster count, $S_i$ is reported as success and removed from $\mathcal{S}$ so that it would not appear afterward.
If the point size reaches 0px or 20px, $S_i$ is marked as a failure and also removed from $\mathcal{S}$.
This adjustment procedure was established to verify whether point size can offset the perceptual bias made by the alter of scatterplot size.

There was no wrap-up interview for this experiment.

\noindent
\textbf{Participants and Appratus.}
We recruited 20 new participants (aged from 21 to 35 years, 24$\pm$ 2.1). The participants' experience with analyzing scatterplots varied: 11 had some experience, 7 had extensive experience, and 2 had little or no experience. The viewing conditions and apparatus remained consistent with those used in the preliminary experiment.

\subsubsection{Results and Discussions}

\begin{figure}[t]
  \centering
  
    \includegraphics[width=\linewidth]{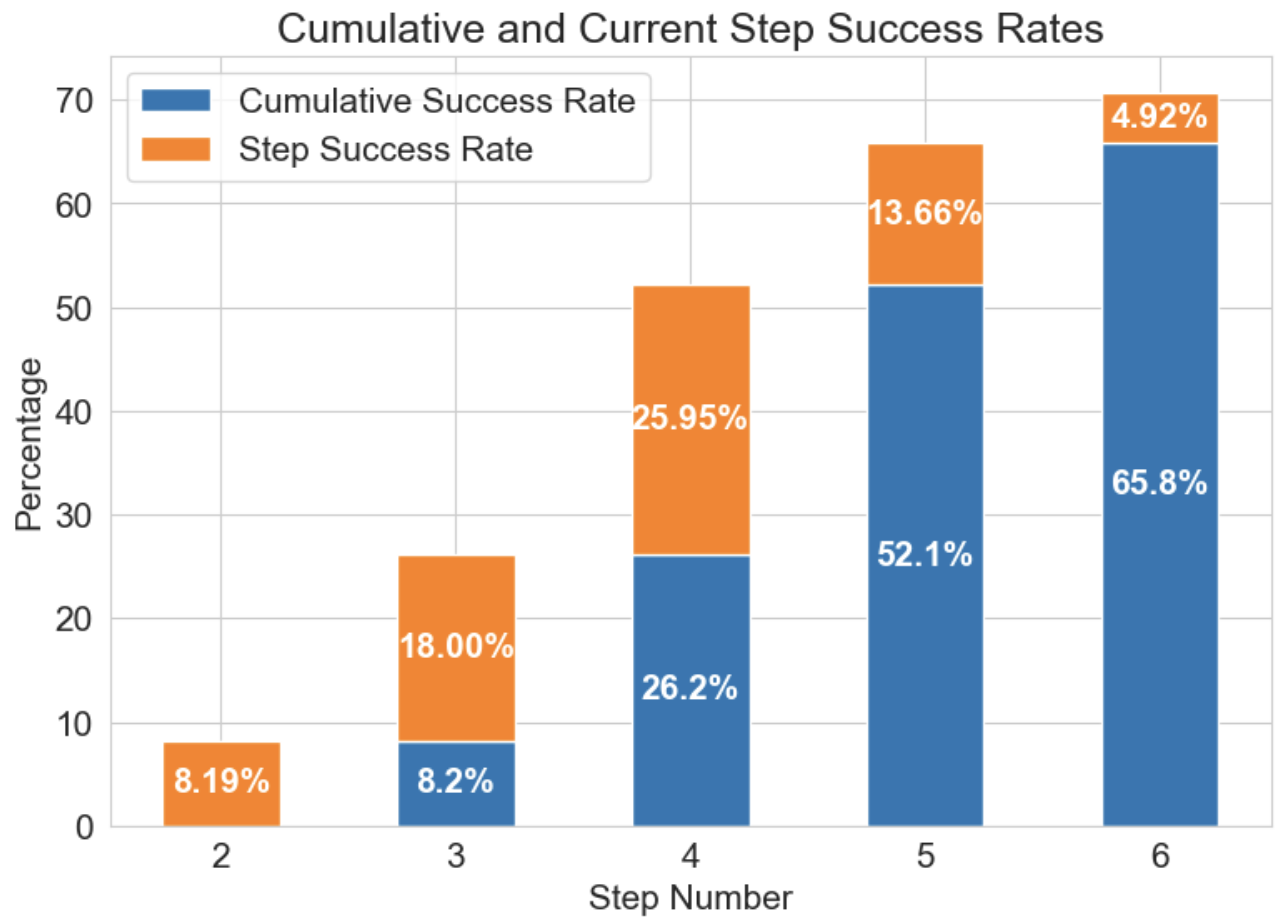}
    \caption{The cumulative and individual step success rates in our main experiment. An increase from each step number indicates that the point change successfully offset the perceptual bias resulting from changes in scatterplot size.}
    \label{fig:cumulative}
\end{figure}

We first investigated how the success rate increased by step number. 
To do so, we first removed 34 cases where the number of clusters did not change between step 0 and step 1. 
We then computed the individual and cumulative success rates of each step. As a result (\autoref{fig:cumulative}), we found that the success rate gradually increased, resulting in 70.7\% at last (Step 6). 
However, the increase eventually slows down between Steps 5 and 6.

\begin{figure}[t]
  
    \includegraphics[width=\linewidth]{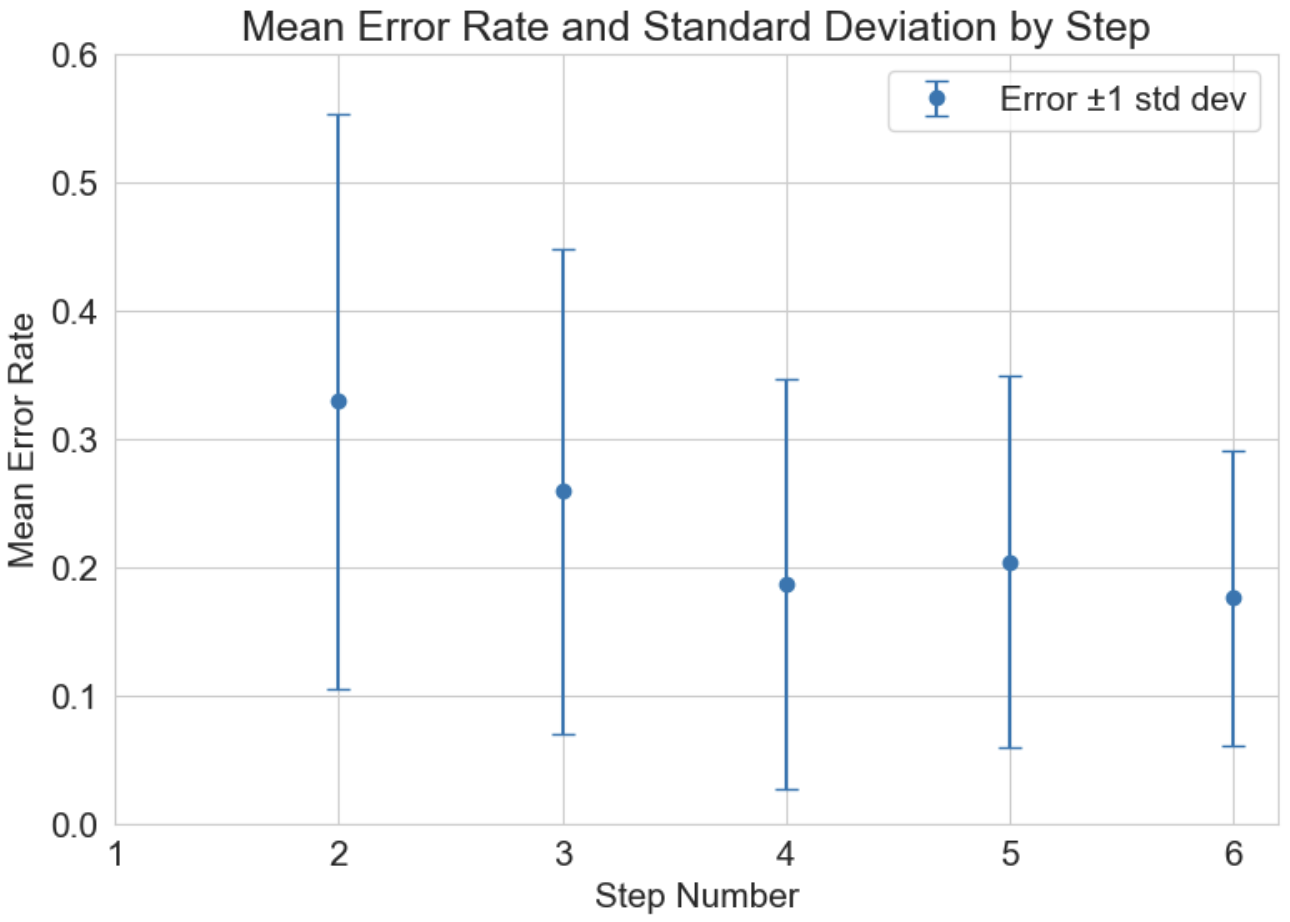}
    \caption{Mean error rate for failed attempts to offset perceptual bias in our main experiment. The decrease in mean error rates indicates that the perceived cluster count progressively converged towards the initial cluster count. The line denotes the 95\% confidence interval.}
    \label{fig:errorrate}

\end{figure}

Then, we calculated the mean error rate at each step.
Formally, the mean error rate is defined as:
\[
\text{Mean Error Rate} = \frac{\left| \text{Initial Cluster \#} - \text{Perceived Cluster \#} \right|}{\text{Initial Cluster \#}}.
\]
As seen in \autoref{fig:errorrate}, mean error rates gradually decrease as the steps in increase from 2 to 6.

Such results clearly indicate that adjusting \textit{point size can mitigate perceptual bias on visual clustering} occurred due to scatterplot size changes. However, the fact that the increment of success rate becomes gradual as step number increases informs that there still remains bias that can hardly be offsetted by just changing point size.

%  We collected cases when point size changes between step 2 and step 6 successfully led participants to view the same number of clusters as seen from Step 0. 259 datasets out of 366 scatterplots (70.7\%) accurately clustered within 5 point size changes. 

% The mean error is calculated at each step as the absolute difference between the initial cluster count and the perceived cluster count, normalized by the initial cluster count. This is given by the equation:
% \[
% \text{Mean Error Rate} = \frac{\left| \text{Initial Cluster} - \text{Perceived Cluster} \right|}{\text{Initial Cluster}}
% \]
% We present successrate at each step at figure # and Mean error rates at figure \#
 
% In figure , we can observe that the  and standard deviation also decreasing. 

% Discussion.tex

\section{Discussions}

In this research, we found that scatterplot size can alter cluster perception and verified that adjusting point sizes can mitigate such perceptual bias. Our limitation is that we lack a grounded, automatic methodology that guides us in determining the most appropriate scatterplot and point sizes.

To resolve this remaining challenge, we further aim to develop a mathematical model that can predict the most appropriate point size for a given scatterplot pattern and size. To this end, we aim to conduct a large-scale study that can cover diverse cluster features (e.g., data distribution, number of clusters, density). 

We believe that the model will substantially aid visual analytics by enabling the following applications:

\noindent
\textbf{Communicative/collaborative analytics environments.}
In communicative and collaborative environments, people may see the same scatterplot with different devices.
We can think about a situation in which the lecturer is explaining data science techniques using scatterplots (\autoref{fig:smallscatter}).
By utilizing the model to adjust scatterplot size and point size, 
we can make cluster perception consistent across different analysts.

\noindent
\textbf{Responsive visualization.}
We can also apply the model to make visualizations responsively adjust their visual encoding, reacting to the changes in display (e.g., foldable devices \cite{schirra21imx}) or window size.
This will help analysts or stakeholders to consistently maintain their perception of the scatterplots, thus potentially reducing confusion while making decisions based on the scatterplots.

% By linking scatterplot size and point size adjustments directly, the model facilitates a clearer understanding

% \subsubsection{Classroom Use}
% One practical application of this model is in educational settings, particularly in data science and statistics classrooms. Educators can employ this model to demonstrate to students how visual scale affects data interpretation. By linking scatterplot size and point size adjustments directly, the model facilitates a clearer understanding of cluster analysis, helping students visually grasp complex concepts in data clustering without misinterpretation due to scaling issues.

% \subsubsection{Parameter Linking in Software}
% Another significant application is in the development of data visualization software, where our model can be integrated to automatically adjust point sizes in response to changes in scatterplot dimensions. This feature would be particularly useful in analytical tools used across various industries, ensuring that insights derived from visual data are consistent and reliable, regardless of the display size or user interface adjustments.

% \subsection{Limitation}

    % Conclusion.tex
\section{Conclusion}

Data analysis can be conducted using various displays, such as smartphones, desktop computers, and tabletop displays. Consequently, the size of scatterplots may vary due to these differences.
In this research, we reveal that such changes in scatterplot sizes can alter visual clustering and verify that adjusting point sizes can mitigate such perceptual bias. 
Our work opens up the opportunity to build automated systems that adjust the visual encoding of scatterplots in various displays to make visual perception consistent over time and space.

\bibliographystyle{abbrv-doi}

\bibliography{template}
\end{document}